\documentclass[a4paper]{jpconf}
\usepackage{graphicx}
\begin{document}
\title{Is there a relativistic nonlinear generalization of quantum mechanics?}

\author{Hans-Thomas Elze}

\address{Dipartimento di Fisica ``Enrico Fermi'' \\ 
        Largo Pontecorvo 3, I-56127 Pisa, Italia }

\ead{elze@df.unipi.it}

\begin{abstract}
Yes, there is. -- A new kind of gauge theory is introduced, where the minimal 
coupling and corresponding covariant derivatives are defined in the space of functions 
pertaining to the functional Schr\"odinger picture of a given field theory. 
While, for simplicity, we study the example of a ${\cal U}(1)$ symmetry, 
this kind of gauge theory can accommodate other symmetries as well. 
We consider the resulting relativistic nonlinear extension of quantum mechanics and 
show that it incorporates gravity in the (0+1)-dimensional limit, where it leads to  
the Schr\"odinger-Newton equations. Gravity is encoded here into a universal nonlinear extension of quantum theory. The probabilistic interpretation, i.e. Born's rule, holds  
provided the underlying model has only dimensionless parameters.
\end{abstract}
%%%%%%%%%%%%%%%%%%%%%%
\section{Introduction}
Linearity of the (functional) Schr\"odinger equation and the validity 
of the superposition principle have been essential ingredients of quantum 
(field) theory since its earliest days. Practically all  
physical phenomena behave nonlinearly when examined 
over a sufficiently large range of the dynamical parameters that determine their  
evolution. What singles out the linear dynamics for the wave function(al)?
Quantum mechanics has been tested experimentally under a 
wide range of laboratory conditions and confirmed in all known cases. 
Yet the mathematical structure of the 
theory, so far, hinges heavily on its linearity embodied in linear 
operators acting on states represented by rays in a Hilbert space 
\cite{vonNeumann55,BZW71}.   

This raises the question: Are nonlinear extensions 
possible which agree with the standard formulation in its experimentally ascertained 
domain of validity? 

If so, could this alleviate the unresolved measurement problem 
\cite{vonNeumann55,BZW71,Adler03}? While the outcome of this second question 
is still open, 
it seems worth while to mention that in recent studies of the related 
wave function collapse or reduction mechanisms by Pearle \cite{Pearle} and 
by Bassi \cite{Bassi} the authors indicate that a nonlinear 
extension of quantum theory, possibly involving additional degrees of 
freedom, might ultimately account consistently for these effects. 

Our present     
aim is to report on a nonlinear extension of quantum field theory 
based on a new {\it functional gauge symmetry}, which operates 
on the space of field configurations rather than on the underlying spacetime \cite{I06}. 
In particular, we will argue that this theory essentially 
incorporates Newtonian gravity, which invites deliberation 
whether such an approach could be of wider use. 
Gravity, in this picture, appears as 
a manifestation of the nonlinearity of quantum mechanics. 
  
Among the numerous earlier works that have attempted to extend quantum theory in a nonlinear 
way, there are: The work by Kibble and by Kibble and Randjbar-Daemi is 
close to ours in that they consider how nonlinear modifications of quantum field theory 
can be made compatible with Lorentz or more generally coordinate invariance 
\cite{Kibble78,Kibble80}. Besides considering a coupling of 
quantum fields to classical gravity according to general relativity, which induces an intrinsic nonlinearity \cite{Kibble80,Kiefer04}, these authors study mean-field type 
nonlinearities, where parameters of the model are state dependent through their 
assumed dependence on expectations of certain operators. Work by 
Bialynicki-Birula and Mycielski introduces a logarithmic nonlinearity into the nonrelativistic 
Schr\"odinger equation, with which many of the features of standard 
quantum mechanics are left intact \cite{Iwo76}. A number of different nonrelativistic models 
of this kind have been systematically studied by Weinberg, offering also an assessment of 
the observational limits on such modifications of the Schr\"odinger equation 
\cite{Weinberg89}. 
  
Independently, Doebner and Goldin and collaborators have also studied nonlinear modifications 
of the nonrelativistic Schr\"odinger equation \cite{DoebnerGoldin}. 
This was originally 
motivated by attempts to incorporate dissipative effects. Later, however, 
they have shown that classes  
of nonlinear Schr\"odinger equations, including many of those considered earlier, 
can be obtained through nonlinear (in the wave function) transformations of the linear quantum 
mechanical equation. They coined the name ``gauge transformations of the third kind'' in this 
context, in analogy with gauge transformations of the second kind 
(corresponding to the usual minimal coupling). -- In distinction, our functional 
gauge transformations work on the configuration 
space over which the {\it wave functional} is defined. This can be clearly seen  
in the way we introduce {\it covariant} functional derivatives  
(cf. Eqs.\,(\ref{dtcov})--(\ref{dxcov}) in 
Section\,3). (Of course, the functional derivatives 
here are to the functional Schr\"odinger picture of 
quantum field theory -- reviewed (and 
generalized for fermions) in \cite{Jackiw} --   
what ordinary derivatives are to quantum mechanics.)    

The necessity of generalizing quantum dynamics for quantum gravity has been discussed 
in view of the ``problem of time'' and the Wheeler-DeWitt equation 
by Kiefer and by Barbour \cite{Kiefer93,Barbour93}. -- Note that this equation, 
playing the role of the Schr\"odinger equation there, is of 
the form of a constraint operator, i.e. the Hamiltonian of canonical gravity, 
acting on the wave functional, $\hat H\Psi =0$. Two unpleasant features are incorporated here: 
no time derivative appears \cite{Kiefer04,Kiefer93} and, since $\hat H$ is hermitean, nothing 
indicates the possibility of complex solutions \cite{Barbour93}. -- 
Both authors pointed out that nonlinear 
modifications could be a wellcome remedy and in Ref.\,\cite{Kiefer93} it was proposed that 
these might arise as a ``supergauge potential'' defined on configuration space. 
While formally analogous to the gauge connection in our covariant derivatives, however, only 
a preliminary interpretation in terms of certain quantum (vacuum) effects has been given. 

Instead, based on the new functional gauge symmetry, all dynamical and 
constraint equations here will be derived from  
a gauge and Lorentz invariant action. {\it A priori} this  
has nothing to do with gravity, in particular, but may be applied to any  
quantum field theory. 

The importance of a probabilistic 
interpretation of the wave function (``Born rule'') is emphasized in all 
previous works. We will recover this as well.  
However, no understanding of the origins of the proposed nonlinearities has been provided 
before, except in the case of semiclassical gravity studied by Kibble and Randjbar-Daemi 
\cite{Kibble80}.  
This is achieved by our gauge principle, which, surprisingly, does 
incorporate a Newtonian form of gravity, see Section\,6. 

Part of the motivation for the present work comes from recent considerations of a 
possible deterministic foundation of quantum mechanics, as already  
verified in a number of models 
\cite{tHooft01,tHooft03,tHooft06,I05susy,Vitiello01,ES02,Blasone04,Smolin,Adler,I05lambda}. 
General principles and physical mechanisms ruling the construction of a deterministic 
classical model underlying a given quantum field theory are hard to come by, cf. 
Ref.\,\cite{tHooft06}. However, the known toy models are promising, amounting to an existence 
proof -- the quantum harmonic oscillator, for example, can be understood completely in classical 
deterministic terms, see Refs.\,\cite{tHooft01,tHooft03,ES02}. 

We expect that 
with better understanding of the emergence of quantum mechanics also resulting nonlinear 
corrections to quantum mechanics should become visible. Models  
that are based on linear (in the wave function) evolution equations alone presumably are 
not sufficient. Nonlinearity seems essential to go beyond 
the canonical framework of quantum theory. It is a central aspect in the following.    

The paper is organized as follows. 
In Section\,2, we recapitulate the work of Ref.\,\cite{Kibble80}, in order to argue 
that the gauge invariant (quantum) action introduced in Section\,3 
is Lorentz invariant, despite the presence of a fundamental length 
parameter. In Section\,4, dynamical and constraint equations are presented and 
the crucial ``nonlinearity factor'' of the action is determined. In Section\,5, we discuss  
the validity of the Born rule in the resulting nonlinear quantum theory. 
In the onedimensional limit, considering stationary states, 
it leads to the Schr\"odinger-Newton equations, see Section\,6. 
Section\,7 presents our concluding remarks.      

%%%%%%%%%%%%%%%%%%%%%%%%%%%%%%%%%%%%%%%%%%%%%%%%%%
\section{Space-time and the Schr\"odinger picture}
We briefly recall here the work of Kibble and Randjbar-Daemi \cite{Kibble80}.  
Consider a four-dimensional globally hyperbolic manifold ${\cal M}$ with a 
{\it given} metric $g_{\mu\nu}$ of signature $(1,-1,-1,-1)$.\footnote{We use units 
such that $\hbar =c=1$.} Then, it is  
possible to globally slice space-time into space-like hypersurfaces, 
such that a chosen family of such surfaces, $\{\sigma (t)\}$, is locally determined 
by: 
\begin{equation}\label{xmu}
x^\mu =x^\mu (\xi^1,\xi^2,\xi^3;t) 
\;\;, \end{equation} 
in terms of intrinsic coordinates $\xi^r$, and there exists an everywhere 
time-like vectorfield $n^\mu$, the normal, with $n_\mu n^\mu =1$ and 
$n_\mu x^\mu_{,r}=0$, where $x^\mu_{,r}\equiv\partial x^\mu/\partial\xi^r$. 
We will need the derivative with respect to $t$ at fixed $\xi^r$ of a  
function $f$, $\dot f\equiv\partial f/\partial t\vert_{\xi}$. In particular, 
then, the lapse function $N$ and shift vector $N^r$ are introduced by  
$\dot x^\mu =Nn^\mu +N^rx^\mu_{,r}$\,, the geometrical meaning 
of which is illustrated, for example, in Chapter\,3.3 of reference \cite{Kiefer04}. 

We begin with a given Lagrangean $L$ of a field theory, 
such as for a real scalar field $\phi$: 
\begin{equation}\label{Lagrangean} 
L\equiv\frac{1}{2}g^{\mu\nu}\partial_\mu\phi\partial_\nu\phi -V(\phi )
\;\;, \end{equation} 
where $V(\phi )$ incorporates mass or selfinteraction terms.  
The corresponding invariant action is: 
\begin{equation}\label{action} 
S\equiv\int\mbox{d}^4x\;\sqrt{-g}L 
\;\;, \end{equation} 
where $g\equiv\det g_{\mu\nu}$. This, in turn, yields the stress-energy tensor $T^{\mu\nu}$:   
\begin{equation}\label{Tmunu} 
\frac{1}{2}\sqrt{-g}T^{\mu\nu}\equiv\frac{\delta S}{\delta g_{\mu\nu}} 
=\frac{1}{2}\sqrt{-g}\left (\partial^\mu\phi\partial^\nu\phi -g^{\mu\nu}L\right )
\;\;. \end{equation} 
With the help of the induced metric $\gamma_{rs}$ on $\sigma (t)$, 
$\gamma_{rs}\equiv g_{\mu\nu}x^\mu_{,r}x^\nu_{,s}$, and the hypersurface 
element $\mbox{d}\sigma_\mu\equiv\mbox{d}^3\xi\;\sqrt{-\gamma}n_\mu$,  
the surface-dependent Hamiltonian can be defined: 
\begin{equation}\label{Hamiltonian} 
H(t)\equiv\int_{\sigma (t)}\mbox{d}\sigma_\mu T^\mu_\nu\dot x^\nu 
\;\;. \end{equation} 
In the simplest case, with $\dot x^\mu =\delta^\mu_0$ (i.e., $N=1$, $N^r=0$) and 
$x^\mu_{,r}=\delta^\mu_r$, these relations become $\gamma_{rs}=g_{rs}$ and 
$H(t)=\int_{\sigma (t)}\mbox{d}^3\xi\;T^{00}$, as expected.  
  
If the stress-energy tensor can be expressed in terms of 
canonical coordinates and momenta, say, the scalar field $\phi =\phi (\xi^i,t)$ and 
its conjugated momentum $\Pi =\Pi (\xi^i,t)$ on time slices $\sigma (t)$, we may assume that the corresponding quantized theory exists, with $\phi$ and $\Pi$ fullfilling standard    
equal-$t$ commutation relations. 
Matters are not that simple in a general curved background.  
Therefore, a heuristic 
derivation of the Schr\"odinger picture from the manifestly covariant Heisenberg 
picture has been presented in Ref.\,\cite{Kibble80}. We will not pursue this,   
since our aim is only to recover their Lorentz invariant form of 
the functional Schr\"odinger equation, a generalization of which will follow 
from the action principle of the following 
section.\footnote{As remarked in Ref.\,\cite{Kibble80}, 
the derivation from an action principle guarantees the general coordinate 
invariance of the theory. However, the 
Schr\"odinger picture clearly depends on the slicing of space-time as well 
as on the parametrisation of the slices. Thus,   
invariance under surface deformations 
-- which can be restricted to diffeomorphism invariance \cite{Kiefer04} -- 
is not implied.}    

In fact, the functional Sch\"odinger 
equation obtained by Kibble and Randjbar-Daemi appears naturally 
as one would guess: 
\begin{equation}\label{Schroedinger} 
i\dot\Psi =H(t)\Psi 
\;\;. \end{equation} 
Using the surface element $\mbox{d}\sigma_\mu$ given above, together 
with Eq.\,(\ref{Hamiltonian}), and: 
\begin{equation}\label{tDerivLocal}
\dot\Psi = \int_{\sigma (t)}\mbox{d}^3\xi\;\dot x^\mu\frac{\delta}{\delta x^\mu}\Psi
\;\;, \end{equation} 
the Schr\"odinger equation can also be 
represented in a local form: 
\begin{equation}\label{SchroedingerLocal} 
i\frac{\delta}{\delta x^\mu}\Psi =\sqrt{-\gamma}n_\nu T^\nu_\mu\Psi
\;\;. \end{equation}  
Thus, the functional Schr\"odinger equation 
can be written in a way that makes explicit the behaviour under Lorentz transformations. 
We specialize to the case of a flat background space-time in the following, 
where field quantization is well understood.    

%%%%%%%%%%%%%%%%%%%%%%%%%%%%%%%%%%%%%%%%%%%%%%%
\section{A new action for a new gauge symmetry}
We consider the generic scalar field theory described by the Lagrangean of 
Eq.\,(\ref{Lagrangean}), 
while internal symmetries and fermions can be introduced as we discussed earlier 
in the second of Refs.\,\cite{I06}.  
Furthermore, specializing the result of the previous section for {\it Minkowski space}, 
we find: 
\begin{equation}\label{HamiltonianMinkowski} 
H(t)=\int_{\sigma (t)}\mbox{d}^3\xi\;T^{00}
=\int\mbox{d}^3x\;\Big\{ -\frac{1}{2}\frac{\delta^2}{\delta\phi^2}
+\frac{1}{2}(\nabla\phi )^2+V(\phi )\Big\}
\equiv H[\hat\pi ,\phi]
\;\;, \end{equation} 
i.e., the usual Hamiltonian which is independent of the parameter time $t$;   
intrinsic and Minkowski space coordinates have been identified, $\vec\xi =\vec x$. 
 -- Here,  
quantization is implemented by substituting the canonical momentum $\pi$ 
conjugate to the field $\phi$ (i.e. the ``coordinate''): 
\begin{equation}\label{momentum}
\pi (\vec x)\;\longrightarrow\;\hat\pi (\vec x)\equiv
\frac{1}{i}\frac{\delta}{\delta\phi (\vec x)}
\;\;. \end{equation}
Correspondingly, we have $\Psi =\Psi [\phi ;t]$, i.e. 
a time dependent functional, in this coordinate representation, and $\dot\Psi =\partial_t\Psi$. 
So far, this is the usual functional Schr\"odinger picture of quantum field theory 
applied to the example of a scalar model \cite{Jackiw,JackiwKerman}.  

Next, we introduce {\it functional gauge transformations} \cite{I06}:  
\begin{equation}\label{localfuncgt} 
\Psi'[\phi ;t]=\exp (i\Lambda [\phi ;t])\Psi [\phi ;t]
\;\;, \end{equation} 
where $\Lambda$ denotes a time dependent real functional. These ${\cal U}$(1) 
transformations are local in the space of field configurations.   
They differ from the usual    
gauge transformations in QFT, since we 
introduce covariant derivatives by the following replacements:  
\begin{eqnarray}\label{dtcov}
\partial_t&\longrightarrow &{\cal D}_t\equiv
\partial_t-i{\cal A}_t[\phi ;t]
\;\;, \\ \label{dxcov} 
\frac{\delta}{\delta\phi (\vec x)}&\longrightarrow &
{\cal D}_{\phi (\vec x)}\equiv
\frac{\delta}{\delta\phi (\vec x)}-i{\cal A}_\phi [\phi ;t,\vec x]
\;\;. \end{eqnarray} 
The real functional ${\cal A}$ presents a new kind of `potential' or `connection'. 
Generally, ${\cal A}$ depends on $t$. However, 
it is a {\it functional} of $\phi$ in Eq.\,(\ref{dtcov}), while it is a 
{\it functional field} in Eq.\,(\ref{dxcov}). 
We distinguish these components of ${\cal A}$ by the subscripts. 
They are required to transform as: 
\begin{eqnarray}\label{Afunctionalgt} 
{\cal A}'_t[\phi ;t]&=&{\cal A}_t[\phi ;t]+
\partial_t\Lambda [\phi ;t]
\;\;, \\ \label{Afunctiongt}  
{\cal A}'_\phi [\phi ;t,\vec x]&=&
{\cal A}_\phi [\phi ;t,\vec x]+
\frac{\delta}{\delta\phi (\vec x)}\Lambda [\phi ;t]
\;\;. \end{eqnarray} 
Applying Eqs.\,(\ref{localfuncgt})--(\ref{Afunctiongt}), it follows 
that the correspondingly generalized functional Schr\"odinger equation 
is invariant under the ${\cal U}$(1) gauge transformations. 

Furthermore, it is suggestive to introduce an invariant `field strength':
\begin{equation}\label{field} 
{\cal F}_{t\phi}[\phi ;t,\vec x]\equiv 
\partial_t{\cal A}_\phi [\phi ;t,\vec x]
-\frac{\delta}{\delta\phi (\vec x)}
{\cal A}_t[\phi ;t]
\;\;, \end{equation} 
in close analogy to ordinary gauge theories; 
note that ${\cal F}_{t\phi}=[{\cal D}_t,{\cal D}_\phi]/(-i)$. 
     
A consistent dynamics for the gauge `potential' 
${\cal A}$ has to be postulated, in order to give a 
meaning to the above `minimal coupling' 
prescription. All elementary fields are   
present as the coordinates on which the wave 
functional depends -- presently just a scalar field, besides time.  
We consider the following ${\cal U}$(1) invariant action:  
\begin{equation}\label{Action}
\Gamma\equiv\int\mbox{d}t\mbox{D}\phi\;\Big\{ 
\Psi^*\Big ({\cal N}(\rho )
\stackrel{\leftrightarrow}{i{\cal D}}_t
-H[\frac{1}{i}{\cal D}_{\phi},\phi ]\Big )\Psi 
+\frac{l^2}{2}\int\mbox{d}^3x\;\big ({\cal F}_{t\phi}
\big )^2\Big\}
\;\;, \end{equation} 
where    
$\Psi^*{\cal N}\stackrel{\leftrightarrow}{i{\cal D}}_t\Psi
\equiv\frac{1}{2}{\cal N}
\{\Psi^*i{\cal D}_t\Psi
+(i{\cal D}_t\Psi )^*\Psi\}$, and with   
a dimensionless real function ${\cal N}$ which depends on the density:  
\begin{equation}\label{rho}
\rho [\phi ;t]\equiv\Psi^*[\phi ;t]\Psi [\phi ;t]
\;\;. \end{equation}
The function ${\cal N}$ incorporates a {\it necessary nonlinearity}, which  
will be uniquely determined in Section\,4, cf. Eq.\,(\ref{nonlin}). The fundamental parameter $l$ has dimension $[l]=[length]$, for dimensionless measure $\mbox{D}\phi$ and $\Psi$,  
independently of the dimension of space-time. 

% \footnote{Note that this parameter has \underline{necessarily} 
% the dimension of a length, in order to give the action its correct dimension; it presents the 
% coupling constant of our theory and will be related to Newton's constant in Section\,6. By 
% suitably rescaling the gauge `potentials', the coupling constant could be moved to the covariant 
% derivatives, as originally 
% discussed \cite{I06}; however, as is familiar from ordinary non-Abelian gauge theories, 
% the equivalent action is often more convenient where the coupling appears only in one place.} 

Our action $\Gamma$ generalizes the one 
employed in Dirac's variational principle for QFT, especially for a scalar field,   
in Refs.\,\cite{Kibble80,JackiwKerman}. 
The quadratic part in ${\cal F}_{t\phi}$ is the 
simplest possible extension, i.e. local in  
$\phi$ and quadratic in the derivatives, together with 
the nonlinearity ${\cal N}(\rho )$. 

An immediate consequence of ${\cal U}$(1) invariance is 
that the Hamiltonian $H$, unlike in QFT, 
cannot be arbitrarily shifted by a constant $\Delta E$, gauge transforming  
$\Psi\rightarrow\exp (-i\Delta Et)\Psi$. Thus, there is an absolute 
meaning to the zero of energy in this theory. 

Translation invariance of the action, Eq.\,(\ref{Action}), 
implies a {\it conserved energy functional}, where a contribution 
which is solely due to 
${\cal A}_t$ and ${\cal A}_\phi$ is added to the matter term, 
which is modified by the covariant derivatives. 

According to Section\,2, 
the Lorentz invariance 
of this theory is guaranteed. The action can be written 
in a Lorentz (and Poincar\'e) invariant way, using the appropriate surface-dependent 
Hamiltonian, cf. Eq.\,(\ref{Hamiltonian}), despite that a fundamental length $l$ 
enters.\footnote{The coordinates $x^\mu$, of course, must not be confused with 
the intrinsic coordinates $\xi^i$ and time parameter $t$.} 

The action depends on 
$\Psi ,\Psi^*,{\cal A}_t$, and 
${\cal A}_\phi$. The equations of motion and a constraint 
will be obtained by 
varying $\Gamma$ with respect to these variables. 

%%%%%%%%%%%%%%%%%%%%%%%%%%%%%%%%%%%%%%%%%%%%%%%%%%
\section{The equations of motion and a constraint}
The dynamical equations of motion are reproduced here 
for convenience, which were previously derived in 
Refs.\,\cite{I06}. The gauge covariant equation for the $\Psi$-functional is:   
\begin{equation}\label{ginvfSscalar}
\left (\rho {\cal N}(\rho )\right )' 
i{\cal D}_t\Psi [\phi ;t]
=H[\frac{1}{i}{\cal D}_{\phi},\phi ]
\Psi [\phi ;t]
\;\;, \end{equation}
where $f'(\rho )\equiv\mbox{d}f(\rho )/\mbox{d}\rho$; it replaces 
the usual functional Schr\"odinger equation.   

The nonlinear Eq.\,(\ref{ginvfSscalar}) preserves the normalization of 
$\Psi$. Fixing it at an initial parameter time, 
in terms of an arbitrary constant ${\cal C}_0$:  
\begin{equation}\label{norm}
\langle\Psi |\Psi\rangle\equiv\int\mbox{D}\phi\;\Psi^*\Psi = {\cal C}_0 
\;\;, \end{equation} 
it is conserved under further evolution, while  
the overlap of two different states, $\langle\Psi_1|\Psi_2\rangle$, may 
vary. This is a necessary ingredient of a probability interpretation 
related to $\Psi^*\Psi$, which will be discussed in more detail in the next 
section. 

Completing the dynamical equations, there is an invariant `gauge field equation': 
\begin{equation}\label{fieldeq} 
\partial_t{\cal F}_{t\phi}[\phi;t,\vec x]
=\frac{-1}{2il^2}\left ( 
\Psi^*[\phi;t]{\cal D}_{\phi (\vec x)}\Psi [\phi;t]
-\Psi [\phi;t]({\cal D}_{\phi (\vec x)}\Psi [\phi;t])^*
\right )
\;\;. \end{equation}
However, there is no time derivative acting on the variable 
${\cal A}_t$ in the action. Therefore, it acts as a Lagrange multiplier for a constraint, 
which is the gauge invariant `Gauss' law':
\begin{equation}\label{Gauss}
\int\mbox{d}^3x\;\frac{\delta}{\delta\phi (\vec x)}{\cal F}_{t\phi}[\phi ;t,\vec x]
=\frac{-1}{l^2}\rho {\cal N}(\rho )
\;\;. \end{equation} 
Of course, this differs from QED, for example, and raises the 
question, whether our functional ${\cal U}$(1) gauge symmetry is compatible 
with standard internal symmetries. This is answered  
affirmatively in the second of Refs.\,\cite{I06}. 

The Eq.\,(\ref{Gauss}) can be combined with Eq.\,(\ref{fieldeq}) to result in a 
continuity equation: 
\begin{equation}\label{continuity}
0=\partial_t\Big (\rho{\cal N}(\rho )\Big )
-\frac{1}{2i}\int\mbox{d}^3x\;\frac{\delta}{\delta\phi (\vec x)}
\left ( 
\Psi^*{\cal D}_{\phi (\vec x)}\Psi 
-\Psi ({\cal D}_{\phi (\vec x)}\Psi )^*
\right )
\;, \end{equation}
expressing local ${\cal U}$(1) `charge' conservation in the space of field configurations. 
Functionally integrating Eq.\,(\ref{Gauss}), we find that the 
total `charge' $Q$ has to vanish at all times: 
\begin{equation}\label{charge}
Q(t)\equiv\frac{1}{l^2}\int\mbox{D}\phi\;\rho{\cal N}(\rho )=0
\;\;, \end{equation}    
since the functional integral of a total derivative is zero. 
% This is different from integrating the usual Gauss' law in electrodynamics over all 
% space, for example, were there can be a flux of the fields out to infinity. -- 
The necessity of the nonlinearity now becomes obvious. 
Without it, the vanishing total `charge' could not be implemented, as it 
would be in conflict with the normalization, Eq.\,(\ref{norm}).  

Next, we determine the nonlinearity factor, 
${\cal N}(\rho )\neq 1$.
We would like to implement Eq.\,(\ref{charge}), similarly as the 
normalization, at an initial parameter time $t$. Since it has to be a 
constant of motion, $\partial_tQ(t)=0$, we express this, 
with the help of Eq.\,(\ref{ginvfSscalar}), 
as a condition on $\rho{\cal N}(\rho )$. It is easily seen that the 
only solution here is a linear function: 
\begin{equation}\label{nonlin}
\rho {\cal N}(\rho )={\cal C}_1\Big (\rho -{\cal C}_0(\int\mbox{D}\phi )^{-1}\Big ) 
\;\;, \end{equation}
if one wants to avoid further constraining $\Psi$ or $\Psi^{*}$; 
the latter would make it more difficult, if not impossible, to obtain linear quantum mechanics 
as a limiting case.\footnote{In Ref.\,\cite{I06}, we used a logarithmic function;     
it has to be discarded, since it is not related to a constant of motion.}  

Evidently, the volume of the space of fields, $\Omega\equiv\int\mbox{D}\phi$, needs to be 
regularized, as well as the second functional derivatives at coinciding points 
which appear. A cut-off on field amplitudes has to be introduced together, for 
example, with the point-splitting technique \cite{Jackiw}. A related renormalization 
procedure is an interesting subject for further study, taking into 
account the new functional gauge symmetry.

%%%%%%%%%%%%%%%%%%%%%%%%%%%%%%%%%%%%%%%%%%%%%%%%%%%%%%%%
\section{Interpreting $\Psi^*\Psi$}  
The probability interpretation 
of the density $\rho =\Psi^*\Psi$ (Born's rule) can be applied, if the    
{\it homogeneity property} holds 
\cite{Kibble78,Iwo76,Weinberg89}: 
$\Psi$ and $z\Psi$ ($z\in\mathbf{Z}$) represent the same   
physical state. Thus, states are associated 
with rays in a Hilbert space (instead of vectors). 

In order to investigate the present case, it is useful to consider scale transformations: 
\begin{equation}\label{scale1} 
\rho ={\cal C}_0^{\;a}{\cal C}_1^{-1}\rho'\;\;,\;\;\;
\int\mbox{D}\phi ={\cal C}_0^{1-a}{\cal C}_1\int\mbox{D}\phi'
\;\;, \end{equation} 
such that $\int\mbox{D}\phi'\rho'=1$; we recall that the real measure 
$\mbox{D}\phi$ and constants ${\cal C}_{0,1}$ are chosen dimensionless, without loss 
of generality; $a$ is real. Furthermore, we rescale: 
\begin{equation}\label{scale2}
(\vec x;t)={\cal C}_0^{-a/2}{\cal C}_1^{1/2}(\vec x';t')\;\;,\;\;\; 
(\phi ;{\cal A}_t)={\cal C}_0^{a/2}{\cal C}_1^{-1/2}(\phi';{\cal A}'_t)
\;\;, \end{equation} 
and, consistently: 
\begin{equation}\label{scale3} 
(\delta_\phi ;{\cal A}_\phi ) =
{\cal C}_0^a{\cal C}_1^{-1}(\delta_{\phi'};{\cal A}_{\phi'})  
\;\;. \end{equation}   
Under these transformations, the action transforms as: 
\begin{equation}\label{Actionscale} 
\Gamma ={\cal C}_1\Gamma' 
\;\;, \end{equation}  
where $\Gamma'$ is defined like $\Gamma$, Eq.\,(\ref{Action}), however, 
replacing all quantities by the primed ones.   
To arrive at this result, the Hamiltonian $H$, cf. Eq.\,(\ref{HamiltonianMinkowski}),  
must {\it not contain dimensionfull constants}. 
%, such as in a Lagrangean mass term, $\propto m^2\phi^2$, 
%which could be contained in our model; a selfinteraction of the 
%form $\propto\lambda\phi^4$ introduces a dimensionless coupling $\lambda$ instead.   

There are several implications. -- First, the scale transformations change the overall scale 
of the action, 
say, in units of $\hbar$, by the constant factor ${\cal C}_1$. This is equivalent to the  
rescaling $\hbar =\hbar'/{\cal C}_1$. However, since we prefer to choose units such that 
$\hbar =1$, we should also fix ${\cal C}_1=1$, henceforth. 
%\footnote{The overall sign of 
%$\rho {\cal N}(\rho )$, cf. Eq.\,(\ref{nonlin}), is chosen with hindsight, see Section\,6.} -- Second, since the 
constant ${\cal C}_0$ does not affect the transformation of $\Gamma$, we can always 
choose to normalize the wave functional to ${\cal C}_0=1$, see Eq.\,(\ref{norm}).  

In this way, we see that states, 
as far as $\Psi$ is concerned, are represented by rays. Therefore, 
a probability interpretation of $\Psi^*\Psi$ according to the {\it Born rule} can be maintained.  
This is in agreement with the obervation that Eq.\,(\ref{ginvfSscalar}), if it were 
not for the presence of the covariant derivatives, now appears like the usual 
functional Schr\"odinger equation. Summarizing the previous discussion, we have: 
\begin{eqnarray}\label{nonlin1}
\rho {\cal N}(\rho )&=&\rho -(\int\mbox{D}\phi )^{-1}
\;\;, \\ [1ex] \label{ginvfSscalar1}   
i{\cal D}_t\Psi [\phi ;t]
&=&H[\frac{1}{i}{\cal D}_{\phi},\phi ]
\Psi [\phi ;t]
\;\;. \end{eqnarray} 
However, it must be stressed that the `potentials' ${\cal A}_t$ and ${\cal A}_\phi$ 
are selfconsistently determined through Eqs.\,(\ref{fieldeq})--(\ref{Gauss}). 
Therefore, we arrive here at intrinsically {\it nonlinear quantum 
mechanics}.\footnote{In the second of Refs.\,\cite{I06}, we have argued that 
{\it microcausality} of the present theory holds. -- The {\it weak superposition 
principle} \cite{Iwo76}, generally, must be expected to fail: 
for two non-overlapping sources adding to the right-hand sides of 
Eqs.\,(\ref{fieldeq})--(\ref{Gauss}), the resulting `potentials' must be expected 
to propagate away from the sources in field space. Thus, the sum of two non-overlapping 
solutions $\Psi_{1,2}$ will hardly present a solution of the coupled equations. 
However, {\it if} two {\it stationary non-overlapping solutions} exist, {\it then} their sum 
also presents a solution; see the stationary equations in Section\,6.}     

The difference to standard quantum mechanics also shows up clearly in Eq.\,(\ref{continuity}), 
with the first term now replaced by $\partial_t\rho$: the flux of probability over 
the space of field configurations is affected nonlinearly by $\Psi^*$ and $\Psi$ 
through the `potential' ${\cal A}_\phi$.  

Finally, we remark that {\it in presence of dimensionfull parameters} in the 
Hamiltonian the above scale symmetry, Eqs.\,(\ref{scale1})--(\ref{Actionscale}), 
breaks down. Then, the normalization of $\Psi$ cannot be chosen 
freely, i.e., rays break into inequivalent vectors. In this situation, 
it is appropriate to consider $\Psi$ and 
$\Psi^*$ as giving rise to two oppositely `charged' real components of the wave functional, 
$\Psi_+\equiv (\Psi +\Psi^*)/\sqrt 2$ and $\Psi_-\equiv (\Psi -\Psi^*)/i\sqrt 2$, which 
interact, while preserving the normalization of $\Psi^*\Psi$. Different normalizations, 
then, correspond to physically different sectors of the theory.    

The {\it absence of the homogeneity property} modifies the usual 
measurement theory. In particular, the usual ``reduction of the wave packet'' 
postulate \cite{vonNeumann55} cannot be maintained. This case has been discussed in 
detail in Ref.\,\cite{Kibble78} and formed the starting point for the particular nonlinear theory proposed there, mentioned before in Section\,1. 

%%%%%%%%%%%%%%%%%%%%%%%%%%%%%%%%%%%%%%%%%%%%%%%%%%%%%%%%%%%%%%
\section{Stationary states and Schr\"odinger-Newton equations}
The time dependence in  
Eqs.\,(\ref{ginvfSscalar})--(\ref{Gauss}) can be separated with the Ansatz    
$\Psi [\phi ;t]\equiv\mbox{exp}(-i\omega t)\Psi_\omega [\phi ]$, 
$\omega\in\mathbf{R}$, and consistently 
assuming that the ${\cal A}$-functionals are {\it time independent}. Thus, 
the Eq.\,(\ref{ginvfSscalar}), together with Eq.\,(\ref{nonlin1}), yields: 
\begin{equation}\label{PsiZeroEq} 
\omega\Psi_\omega [\phi ]
=H[\frac{1}{i}{\cal D}_{\phi},\phi ]\Psi_\omega [\phi ]-{\cal A}_t[\phi ]
\Psi_\omega [\phi ]
\;\;, \end{equation}
with ${\cal D}_\phi = \frac{\delta}{\delta\phi}+i{\cal A}_\phi$ and  
$\rho_\omega\equiv\Psi_\omega^*[\phi ]\Psi_\omega [\phi ]$. 
We obtain from Eq.\,(\ref{fieldeq}):  
\begin{equation}\label{fieldeq0} 
\frac{1}{2i}\left ( 
\Psi^*_\omega [\phi ]{\cal D}_{\phi (\vec x)}\Psi_\omega [\phi ]
-\Psi_\omega [\phi ]({\cal D}_{\phi (\vec x)}\Psi_\omega [\phi ])^*
\right )=0
\;\;, \end{equation}
which expresses the vanishing of the `current' in the stationary 
situation. -- 
Applying a time independent gauge transformation, 
cf. Eqs.\,(\ref{localfuncgt}), (\ref{Afunctiongt}), the stationary wave functional 
can be made {\it real}. Then, the Eq.\,(\ref{fieldeq0}) implies ${\cal A}_\phi=0$;   
consequently, ${\cal D}_{\phi}\rightarrow\frac{\delta}{\delta\phi}$ 
everywhere. Finally, `Gauss' law', Eq.\,(\ref{Gauss}), 
determines ${\cal A}_t$: 
\begin{equation}\label{Gauss0}  
\int\mbox{d}^3x\;\frac{\delta^2}{\delta\phi (\vec x)^2}{\cal A}_t[\phi ]
=\frac{1}{l^2}\Big (\rho -(\int\mbox{D}\phi )^{-1}\Big )
\;\;, \end{equation} 
which has to be solved selfconsistently together with Eq.\,(\ref{PsiZeroEq}). -- 
Thus, separation of the time dependence has given us two coupled 
equations. They represent a field theoretic generalization 
of the stationary Schr\"odinger-Newton equations, as we shall now explain. 

The time dependent {\it Schr\"odinger-Newton equations} for a particle of mass $m$ are given by: 
\begin{equation}\label{SN} 
i\hbar\partial_t\psi =-\frac{\hbar^2}{2m}\nabla^2\psi -m\Phi\psi\;\;,\;\;\; 
\nabla^2\Phi =4\pi Gm\vert\psi\vert^2 
\;\;, \end{equation}
where $G\equiv l_P^2c^2/\hbar$ is Newton's gravitational constant 
(here related to the Planck length $l_P$) and $\Phi$ denotes the gravitational 
potential. They represent the nonrelativistic approximation to ``semiclassical gravity'', 
i.e. Einstein's field equations coupled to the expectation value of the operator-valued stress-energy tensor of quantum matter. They are considered in arguments related to 
``semiclassical gravity'', 
to gravitational self-localization of mesoscopic or macroscopic mass distributions, and 
to the role of gravity in the objective reduction scenarios of Di\'osi and of Penrose 
 -- see, for example, the 
Refs.\,\cite{Kibble80,Diosi84,Penrose98,Diosi05,Carlip06,Diosi06}, and further references 
therein.  
  
In a Universe which consists only of a single point, 
our field theory equations (\ref{PsiZeroEq}) and (\ref{Gauss0}) indeed 
reduce to the stationary 
Schr\"odinger-Newton equations {\it in one dimension}. Appropriate rescalings by powers 
of $l$, $m$, $\hbar$, and $c$ of the various quantities have to be incorporated, 
in order to give the 
equations their onedimensional form. 
%, where Newton's constant $G$ is a dimensionless parameter. 
With a nonzero potential $V(\phi )$ in our Hamiltonian, 
the Schr\"odinger equation in (\ref{SN}) would aquire an additional term. 

% --  
% Explicitly, keeping units such that $\hbar =c=1$ and considering the Hamiltonian of 
% Eq.\,(\ref{HamiltonianMinkowski}) with $V(\phi )\equiv 0$, 
% the following substitutions have to be performed, in order to arrive 
% at the stationary limit of Eqs.\,(\ref{SN}): 
% \begin{eqnarray}\label{subs1}
% |\Psi |^2\;\longrightarrow\;4\pi G^2m|\psi |^2&,&\; {\cal A}_t\;\longrightarrow\; -m\Phi\;\;, 
% \\ [1ex] \label{subs2} 
% \int\mbox{d}^3x\;\frac{\delta^2}{\delta \phi (x)^2}\;\longrightarrow\;\frac{1}{m}
% \frac{\mbox{d}^2}{\mbox{d}q^2}&,&\;\int{\cal D}\phi\;\longrightarrow\;(4\pi G^2m)^{-1}
% \int_{-Q/2}^{+Q/2}dq=Q/4\pi G^2m
% \;\;, \end{eqnarray} 
% where $m$ is the relevant (particle) mass scale and Q denotes a regulator length, much larger 
% than any length scale of the one-dimensional system. Of course, the gradient terms of the 
% Hamiltonian, $\propto (\nabla\phi )^2$, do not contribute in this limit (``a single point has 
% no neighbours'').     

More generally, the Schr\"odinger-Newton equations present the formal limit of the present 
gauge theory in 0+1 dimensions, i.e. the quantum mechanical limit related to the usual 
discussions of these equations. This can be generalized 
by considering a (lattice) discretized version of the functional equations, which then  
amounts to a quantum many-body theory incorporating a form of gravity. 

It seems remarkable that the gravitational interaction arises here in the space of 
quantum states (configuration space). Yet, in view of the fundamental length $l$ present 
in the action, Eq.\,(\ref{Action}), it is not a complete surprise that our gauge 
theory incorporates gravity. We notice, however, also a {\it deviation from 
Newtonian gravity}, presented by the constant term on the right-hand side of Eq.\,(\ref{Gauss0}). 
While it is natural to let this term become arbitrarily small in the quantum mechanical limit 
just discussed, its presence is necessary for the full theory, cf. Section\,4.  
% This is an important topic for further study, related to the regularization of the theory. 
Thus, gravity can turn from being 
attractive to being repulsive, depending on whether the right-hand side of 
this equation is negative or positive, respectively. 

In Ref.\,\cite{Carlip06}, it has recently been shown that sufficiently large Gaussian 
wave packets show a tendency to shrink in width as they evolve according to the 
time dependent Schr\"odinger-Newton equations. This leads 
to a {\it decrease of interference effects}, 
which possibly will be observable in near-future molecular 
interference experiments. 
% It will be interesting to study the behaviour of such wave 
% packets according to the present theory. 
% : we expect that the anti-gravitating effect 
% would lead to an even stronger depletion of the low mass density tails of wave packets, 
% {\it enhancing self-localization}. -- 
We speculate that according to the present theory 
coherent superpositions of displaced wave packets 
({\it Schr\"odinger cat states}) decay by giving rise to time dependent 
`potentials' ${\cal A}_t$ and ${\cal A}_\phi$, while attracting each other 
similar to corresponding classical matter distributions. Such behaviour  
could have some impact on dynamical `collapse of the wave function' or reduction theories. 

%%%%%%%%%%%%%%%%%%%%%
\section{Conclusions}
A relativistic ${\cal U}(1)$ gauge theory has been presented which constitutes an 
intrinsically nonlinear extension of quantum mechanics or quantum field theory. 

Closest in spirit is the work of Kibble and Randjbar-Daemi \cite{Kibble80} where 
such nonlinearities -- due to coupling the expectation of the matter 
stress-energy tensor to classical general relativity or due to making parameters of the 
model state dependent -- have been 
discussed in a relativistic setting before. However, this has been reminiscent of a 
mean-field approximation. 
   
In distinction, based on a new gauge principle, we have introduced two `potentials', 
${\cal A}_t$ and ${\cal A}_\phi$, which are {\it not} additional independent fields but 
functionals that depend on the same field variables of the underlying (scalar or other) 
field theory as the wave functional $\Psi$. Their  
dynamical and constraint equations follow from a relativistic invariant 
action principle, introduced in Section\,3. Thus, if the `potentials' 
are eliminated by solving the respective equations, in principle, a nonlinear theory in $\Psi$ 
necessarily results. 

Note that in the absence of quantum matter, 
$\Psi =0$, the Eqs.\,(\ref{fieldeq}) and (\ref{Gauss}) that determine the `field strength' 
${\cal F}_{t\phi}$ -- and similarly in the (0+1)-dimensional limit -- have no time 
dependent solutions. Therefore, the `potentials' do not propagate independently of matter 
sources here.\footnote{This is due to the fact that the analogue of a magnetic 
field is missing for any underlying model based on a one-component field, see  
Eqs.\,(\ref{Lagrangean}) and (\ref{field}). The situation changes in the presence of internal 
symmetries, as discussed in the second of Refs.\,\cite{I06}.} 

We have shown that the homogeneity property holds which is necessary for the  
representation of states by rays in Hilbert space. Thus, the Born rule can be applied, 
giving a probabilistic interpretation to $\Psi^*\Psi$ \cite{Kibble78,Iwo76,Weinberg89}. 
However, it   
breaks down, if the assumed underlying classical model contains dimensionfull parameters. 
In this case, a discussion in terms of the `charged' components of $\Psi$ is appropriate, 
which invites further interpretation.
  
%Since the variables here are the same as in the functional Schr\"odinger picture corresponding 
%to a given classical model, we may assume that the observables of the nonlinear theory 
%are the same as in standard quantum field theory.  

Related to the presence of a fundamental lenght $l$ in the action, in the zerodimensional limit the presented theory recovers the 
Schr\"odinger-Newton equations, coupling Newtonian gravity 
to quantum mechanics \cite{Kibble80,Diosi84,Penrose98,Diosi05,Carlip06,Diosi06}. 
% However, it seems characteristic of the gauge theory (based on the simplest compact 
% group) that gravity 
% changes sign at very low probability densities. In any case, 
Thus, the proposed theory incorporates  
Newtonian gravity into quantum field theory: {\it unlike the standard coupling of 
independent gravitational degrees of freedom to matter, gravity is encoded here into a 
universal nonlinear extension of quantum field theory}.    
%
% In the future, the regularization of the theory and a  
% perturbative scheme need to be worked out, in order to have control of its microscopic 
% behaviour in situations where gravity is weak. 
% It will be interesting whether 
% the presented {\it functional gauge symmetry} 
% can be further generalized and what ensuing experimental predictions of will be. 

\ack   
I am grateful to H.-D.\,Doebner, G.A.\,Goldin, and C.\,Kiefer for discussions 
or correspondence. 

%%%%%%%%%%%%%%%%%%%%%%%%%%%
 
%%%%%%%%%%%%%%

\end{document}